\documentclass[a4paper,11pt]{article}
\pdfoutput=1 

\usepackage{booktabs}
\usepackage{caption}
\usepackage{subcaption}
\usepackage{graphicx}
\usepackage{siunitx}
\usepackage{todonotes}
\usepackage{jinstpub} 
\graphicspath{{figures/}}
\title{\boldmath Silicon Pixel Sensor R\&D for the CLIC Tracking Detector}

\author[a,b,1]{J. Kr\"oger,\note{Corresponding author.}}

\affiliation[a]{University of Heidelberg, Germany}
\affiliation[b]{CERN, Switzerland}

\emailAdd{kroeger@physi.uni-heidelberg.de}

\abstract{
The physics aims at the proposed high-energy $e^+e^-$ collider \textit{CLIC} pose challenging demands on the performance of the detector system.
Precise hit-time tagging, an excellent spatial resolutions, and a low mass are required for the vertex and tracking detectors.
To meet these requirements, an all-silicon vertex and tracking detector system is foreseen, for which a broad R\&D programme on a variety of novel silicon detector technologies is being pursued.
For the ultra-low mass vertex detector, different hybrid technologies with innovative sensor concepts and interconnection techniques are explored.
For the large-scale tracking detector, the focus of the R\&D lies on monolithic HV-MAPS and HR-CMOS technologies.
This contribution gives an overview of the ongoing activities with a focus on monolithic technologies for the CLIC tracking detector. Recent results from laboratory and test-beam measurement campaigns of the \textit{ATLASpix\_Simple} and the \textit{CLICTD} sensor prototypes are presented.
}


\collaboration[c]{on behalf of the CLICdp collaboration}

\proceeding{INSTR20: Instrumentation for Colliding Beam Physics\\
  February 24 - 28, 2020\\
  Nowosibirsk, Russia}

\begin{document}
\maketitle
\flushbottom

\section{Introduction}
\label{sec:intro}
The \textit{Compact Linear Collider} (CLIC)~\cite{CLICCDR_vol1,clic-report} is a proposed high-energy electron-positron collider based at CERN in Geneva, Switzerland, for the era beyond the High-Luminosity Large Hadron Collider (HL-LHC)~\cite{hl-lhc-tdr}.
Its physics goals comprise top quark and Higgs boson measurements with unprecedented precision, as well as searches for Physics Beyond Standard Model (BSM).
It is proposed to be constructed in three energy stages reaching a centre-of-mass energy of up to \SI{3}{TeV} and a total length of \SI{50}{km} at the final stage.

An innovate two-beam acceleration scheme is being developed for the CLIC accelerator~\cite{CLICCDR_vol1}.
A low-energy high-current drive beam transfers its energy to the main beam with high energy and low current in normal-conducting cavities, the so-called two-beam modules.
This acceleration technique allows for gradients of more than \SI{100}{MV/m}.

The proposed physics aims for CLIC~\cite{yellowreport_physics_potential} pose challenging demands on the performance of the detector system~\cite{yellowreport_detectortech}.
For pile-up rejection and to mitigate the impact of beam-induced background, a precise hit-time tagging with a resolution of \SI{\sim 5}{ns} is required for the vertex and tracking detectors.
In addition, an ultra-low material budget of \SI{\sim 0.2}{\percent} of a radiation length ($X_0$) per layer for the vertex detector and \SI{\sim 1}{\percent} $X_0$ per layer for the tracking detector are required while achieving a single-plane spatial resolution of a few micrometers.
The detector requirements are summarised in Table~\ref{tab:detector_requirements}.

\begin{table}[htbp]
\centering
\caption{Summary of the requirements for the CLIC vertex and tracking detectors~\cite{yellowreport_detectortech}.}
\label{tab:detector_requirements}
\begin{tabular}{lcc}
\toprule
& \textbf{Vertex Detector} & \textbf{Tracking Detector} \\
\midrule
\textbf{detector area}				& \SI{0.84}{m\squared} 								& \SI{137}{m\squared} \\
\textbf{timing resolution}			& \SI{\sim5}{ns}									& \SI{\sim5}{ns}\\
\textbf{hit detection efficiency}	& $99.7-99.9$~\SI{}{\percent}						& $99.7-99.9$~\SI{}{\percent} \\
\textbf{maximum pixel size} 		& $(25\times 25)~\SI{}{\micro m \squared}$ 		& $30-50~\SI{}{\micro m}\times 1-10~\SI{}{mm}$ \\
\textbf{material budget per layer} & \SI{\sim 0.2}{\percent}~$X_0$ 					& $1-1.5$~\SI{}{\percent}~$X_0$ \\
\textbf{average power dissipation}	& \SI{< 50}{mW/cm\squared} 							& \SI{< 150}{mW/cm\squared} \\
\bottomrule
\end{tabular}
\end{table}

To meet the stringent detector requirements, an all-silicon vertex and tracking detector system is foreseen as the central part of the \textit{CLIC detector model} (CLICdet)~\cite{clicdet_note}.
In order to identify suitable technologies, a broad R\&D programme on a variety of novel silicon detector technologies is being pursued~\cite{yellowreport_detectortech}.

For the ultra-low mass vertex detector, different small pitch (\SI{25}{\micro m}) hybrid technologies with innovative sensor concepts are explored~\cite{Velyka:2019det}.
A dedicated readout chip called \textit{CLICpix2}~\cite{CLICpix2manual} has been developed in \SI{65}{nm} CMOS technology and bump-bonded to thin planar active-edge sensors~\cite{Williams:2020kbw}.
To overcome the challenges of the fine-pitch bump-bonding, alternative interconnection techniques such as capacitive coupling and anisotropic conductive films are explored~\cite{Spannagel:2020fin}.
In addition, Silicon-On-Insulator (SOI) test chips are also under investigation~\cite{Bugiel:2018ckn}.

Monolithic CMOS technologies are promising for the large-area tracking detector due to their cost effectiveness and large-scale production capabilities.
Different sensors with large and small collection electrodes are under investigation and recent results from the \textit{ATLASpix\_Simple} and the \textit{CLICTD} test chips are presented in Section~\ref{sec:results} of this paper.

In order to predict and further optimise the performance of the various prototype technologies, a fast and versatile simulation tool, \textit{Allpix Squared}~\cite{allpix-squared}, has been developed.
It allows for high statistics Geant4-based~\cite{geant4-reference} Monte Carlo simulations, for which detailed electric field maps from TCAD finite-element simulations can be imported for a highly accurate description of the sensor properties.

To allow fast prototyping cycles, a versatile data acquisition system called \textit{Caribou}~\cite{Vanat:2703500}, has been developed to be used in both laboratory measurements and test-beam campaigns.
It reduces the effort of taking new pixel sensor prototypes into operation by maximizing the fraction of common hardware and software blocks of the system while keeping chip-specific components to a minimum.

In addition, the flexible and modular test-beam data reconstruction framework \textit{Corryvreckan}~\cite{corryvreckan_manual_v1} has been developed.
It fulfils the requirements of a complex offline event building in data taking environments combining detector subsystems with different readout architectures.


\section{Depleted monolithic CMOS Sensors for the CLIC Tracking Detector}
\label{sec:results}
For the tracking detector with its large area of \SI{\sim 140}{m\squared} fully monolithic CMOS technologies are considered the best choice due to their cost efficiency and large-scale production capabilities.
In addition, they allow for a reduced material budget compared to hybrid technologies.

A high voltage CMOS sensor with a large collection diode, the \textit{ATLASpix\_Simple}~\cite{Peric:2019hmv}, has been characterised both in laboratory and test-beam measurements and initial results have been presented in~\cite{yellowreport_detectortech}. Recent results focusing on a comparison of different substrate resistivities are presented in Section~\ref{subsec:atlaspix}.

3D TCAD simulations and previous test results have led to the development of innovative design concepts for CMOS sensors with a small collection electrode~\cite{Munker_2019}.
These have been implemented in various prototype chips targeting both CLIC and other future projects. The \textit{CLICTD} chip~\cite{Kremastiotis:2019pnb} has recently been produced using a modified \SI{180}{nm} CMOS imaging process implemented on a high-resistivity epitaxial layer.
The design includes an innovative sub-pixel segmentation scheme, and first results are presented in Section~\ref{subsec:clictd}.

\subsection{The ATLASpix\_Simple HV-MAPS Prototype}
\label{subsec:atlaspix}
The \textit{ATLASpix\_Simple} sensor is a High-Voltage Monolithic Active Pixel Sensor (HV-MAPS), which was designed as a candidate for the ATLAS ITk upgrade~\cite{Peric:2019hmv} but also targets the requirements of the CLIC tracking detector.

\begin{figure}[b]
\begin{minipage}[b]{.45\linewidth}
\centering
\includegraphics[width=0.5\textwidth]{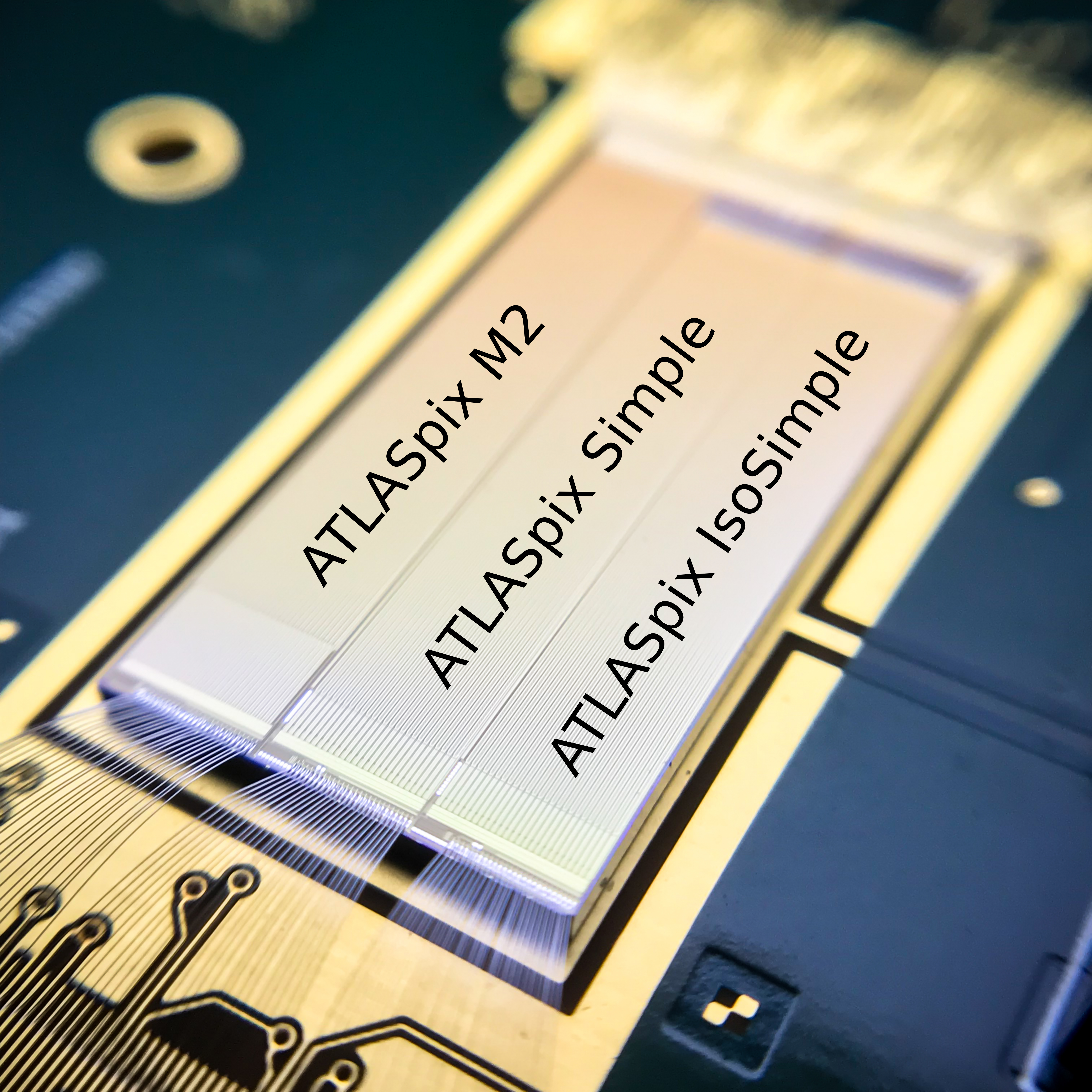}
\caption{Photograph of an \textit{ATLASpix} sensor with its three submatrices glued and wire-bonded to a printed circuit board.}
\label{fig:apx_picture}
\end{minipage}%
\hfill
\begin{minipage}[b]{.45\textwidth}
\centering
\includegraphics[width=\textwidth]{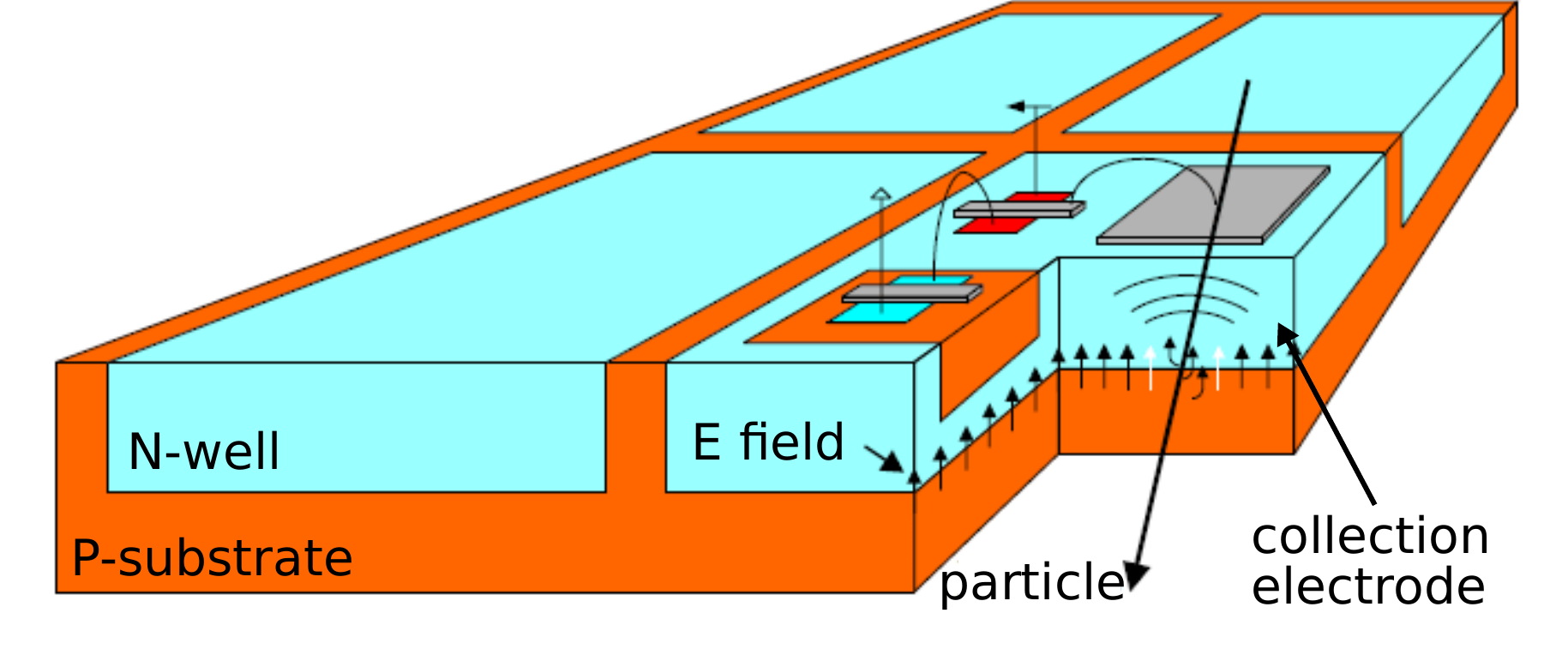}
\caption{Schematic drawing of the \hbox{HV-MAPS} concept~\cite{Peric:2007zz} (modified). \newline}
\label{fig:hv_cmos_sketch}
\end{minipage}
\end{figure}

It was produced in a commercial \SI{180}{nm} HV-CMOS process on wafers with substrate resistivities between 20 and \SI{200}{\ohm cm} and features an active matrix consisting of 25 columns and 400 rows of pixels with a pitch of $130\times40$~\SI{}{\micro m\squared}.
A photograph is shown in Figure~\ref{fig:apx_picture}.

As depicted in Figure~\ref{fig:hv_cmos_sketch}, each pixel consists of a large collection electrode, which is placed in a deep N-well on a p-substrate and houses the in-pixel electronics comprising charge-sensitive amplifier and comparator.
After production, the sensors can be thinned down to \SI{50}{\micro m} by removing undepleted bulk material from the backside.
A high bias voltage of up to $\mathcal{O}$(\SI{100}{V}) leads to a large depleted volume with a high electric field resulting in a fast charge collection via drift.

For each hit, the time-of-arrival (ToA) is recorded with \SI{10}{bit} a precision and a binning of \SI{16}{ns}.
In addition, the signal charge is determined with a time-over-threshold (ToT) measurement with a \SI{6}{bit} precision.
The pixel matrix is read out in a data-driven column-drain scheme.

It has previously been shown that the measured spatial resolution of the \textit{ATLASpix\_Simple} in column and row direction is approximately the pixel pitch$/\sqrt{12}$.
This corresponds to a binary resolution and is in agreement with the fact that almost no charge is shared with neighbouring pixels~\cite{yellowreport_detectortech}.

\subsubsection{Hit Detection Efficiency for Different Substrate resistivities}

\begin{figure}[t]
\centering
\begin{subfigure}[b]{.45\linewidth}
\centering
\includegraphics[width=\textwidth]{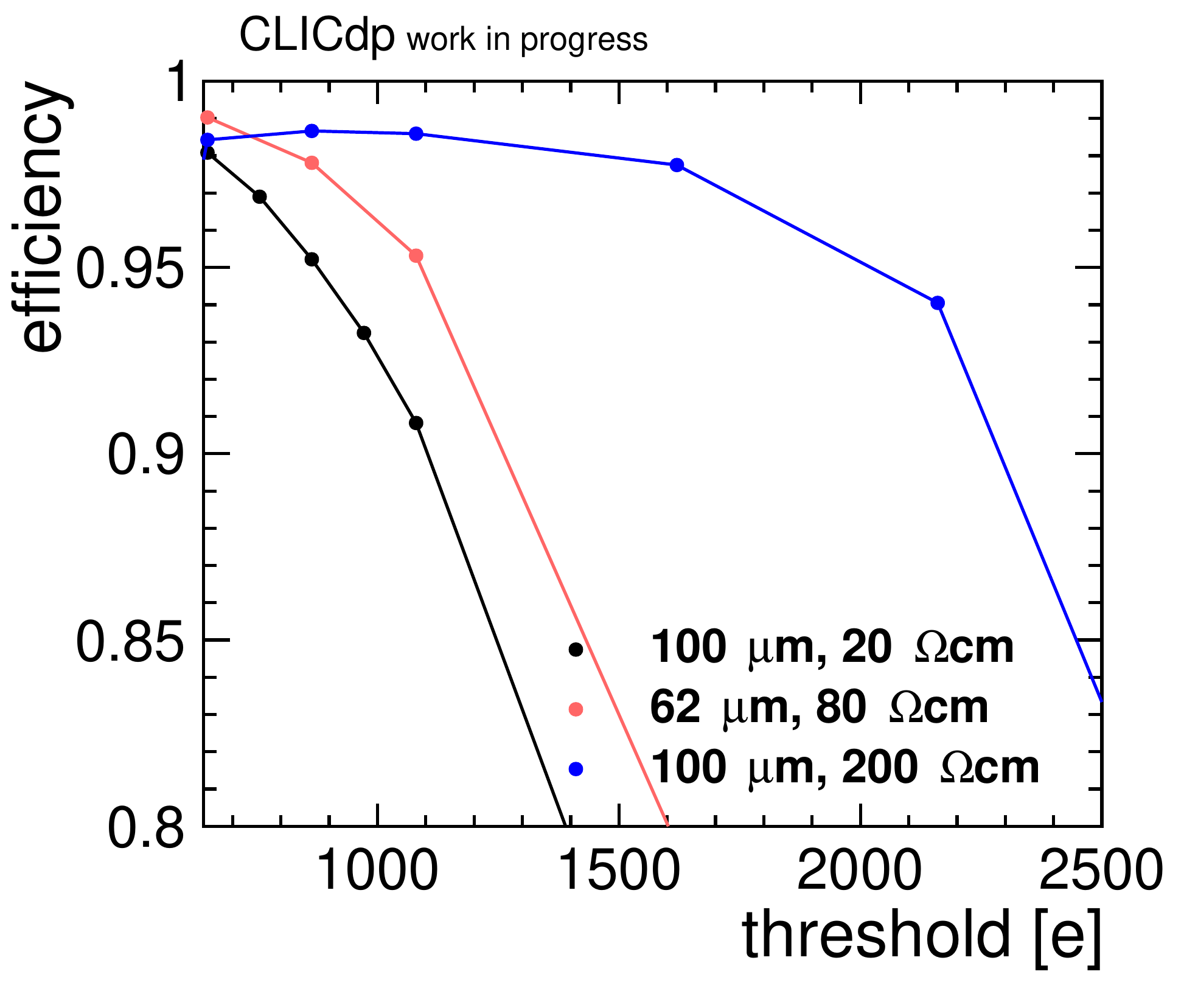}
\caption{Threshold dependence at a bias voltage of \SI{-50}{V}.}
\label{fig:apx_thres_dependence}
\end{subfigure}%
\hfill
\begin{subfigure}[b]{.45\linewidth}
\centering
\includegraphics[width=\textwidth]{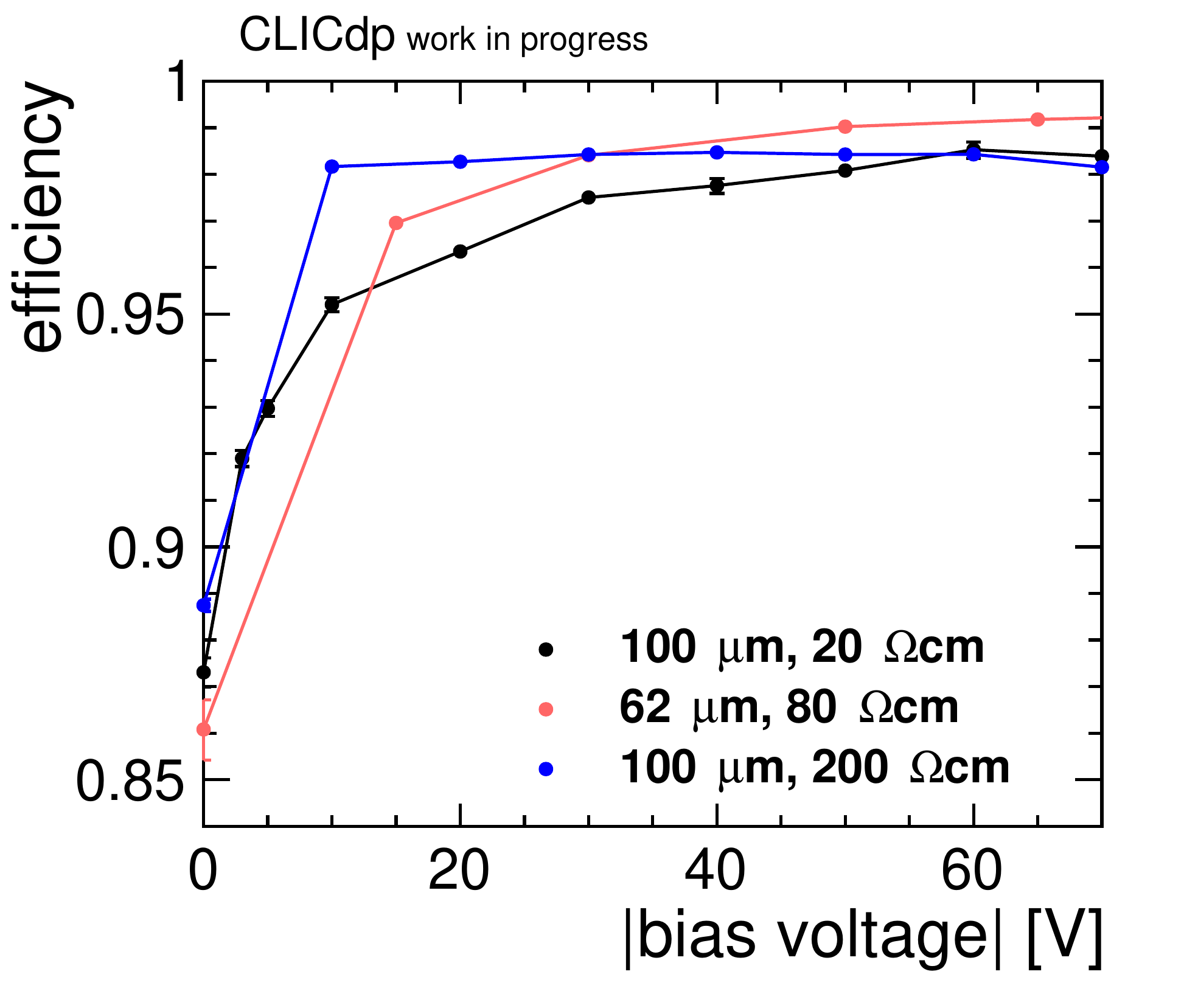}
\caption{Bias voltage dependence at a detection threshold of \SI{650}{e^-}.}
\label{fig:apx_bias_dependence}
\end{subfigure}
\caption{\label{fig:apx_efficiency_thres_and_bias} Hit detection efficiency of \textit{ATLASpix\_Simple} samples with different substrate resistivities and thicknesses.}
\end{figure}

The hit detection efficiency and its dependence on the detection threshold as well as the bias voltage have also been studied in DESY-II test-beam campaigns.
The results are shown in Figure~\ref{fig:apx_efficiency_thres_and_bias}.

As can be seen in Figure~\ref{fig:apx_thres_dependence}, the efficiency drops with an increasing detection threshold.
Comparing the different substrate resistivities, it drops significantly slower for the \SI{200}{\ohm cm} sample than for the lower substrate resistivities.
As a consequence, the \SI{200}{\ohm cm} sample can be operated for a much larger range of the threshold while maintaining a high detection efficiency.

This behaviour is expected because a higher substrate resistivity leads to a larger depleted volume for a given bias voltage as described in Section~\ref{subsec:apx_timing_performance}.
Consequently, the signal induced by the generated charge carriers of an incident particle is larger so that it can exceed the detection threshold for higher thresholds before becoming inefficient.

Figure~\ref{fig:apx_bias_dependence} shows how the hit detection efficiency changes with the applied bias voltage.
It can be observed that the efficiency rises with increasing bias voltage.
For the lower resistivities, the efficiency saturates only slowly whereas it already reaches a stable plateau at a bias voltage around \SI{10}{V} for the \SI{200}{\ohm cm}.
Therefore, the \SI{200}{\ohm cm} sample becomes fully efficient at a much lower bias voltage.

Also this observation meets the expectation since a higher substrate resistivity leads to a larger depleted volume for a given bias voltage.
In return, the \SI{200}{\ohm cm} sample reaches a given depleted volume, and therefore a given signal size, for a smaller applied bias voltage.
Consequently, it becomes fully efficient at a lower bias voltage.

It should be noted that the sensor thickness is not expected to have an influence on the hit detection efficiency because by the thinning only undepleted bulk material is removed from the back of the sensor and the contribution to the charge collection from the undepleted bulk is insignificant.

\subsubsection{Timing Performance for Different Substrate resistivities}
\label{subsec:apx_timing_performance}
The timing performance for the different substrate resistivities of the \textit{ATLASpix\_Simple} has been determined in measurement campaigns at the DESY-II test-beam facility~\cite{desy2_testbeam_facility} using an EUDET-type beam telescope~\cite{Jansen:2016bkd} and an additional Timepix3 plane~\cite{timepix3_paper} as a timing reference.

\begin{figure}[t]
\centering
\begin{subfigure}[b]{.33\linewidth}
\centering
\includegraphics[width=\textwidth]{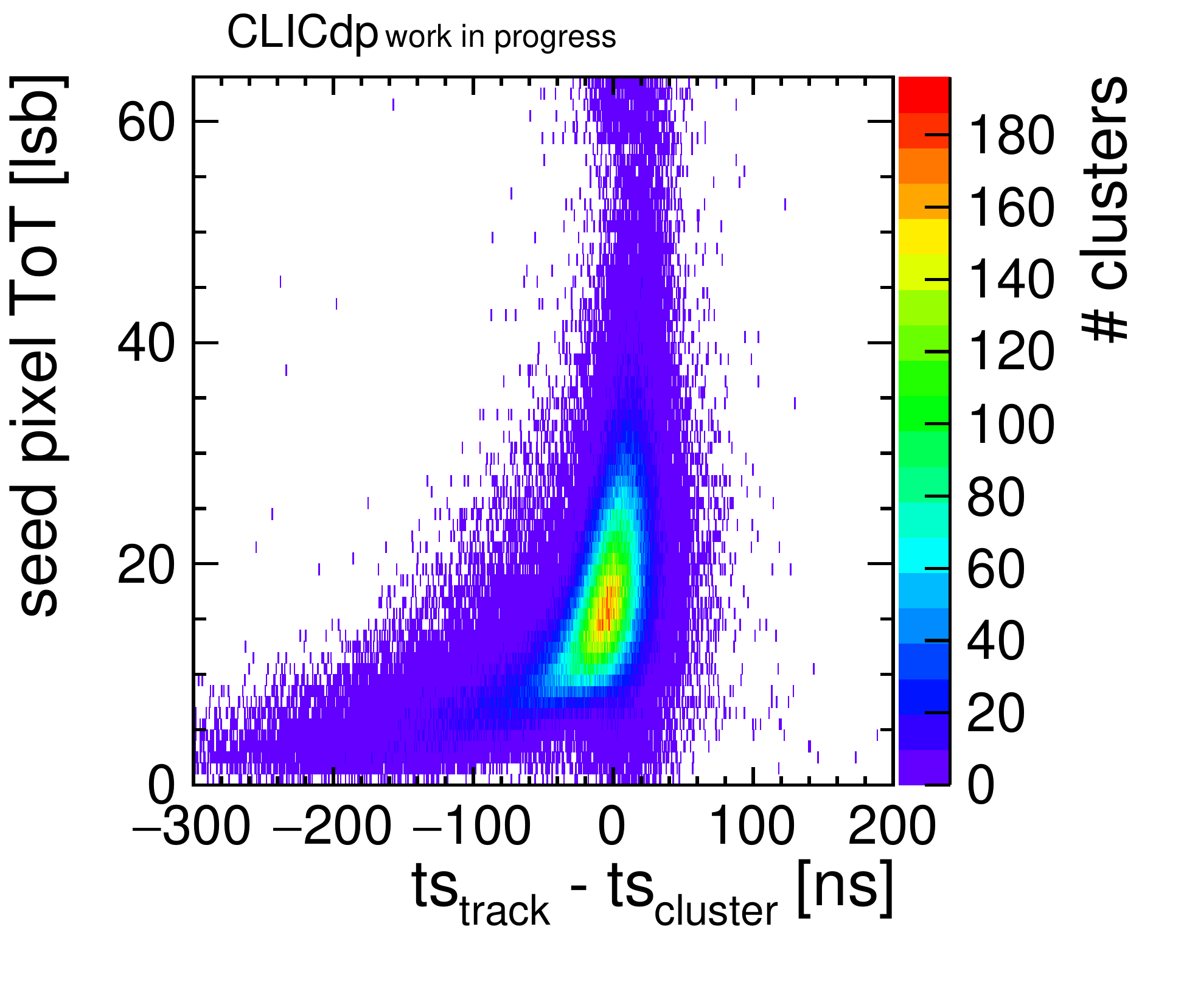}
\caption{\SI{20}{\ohm cm}, \SI{100}{\micro m}.}
\label{fig:apx_compare_timewalk_20}
\end{subfigure}%
\begin{subfigure}[b]{.33\linewidth}
\centering
\includegraphics[width=\textwidth]{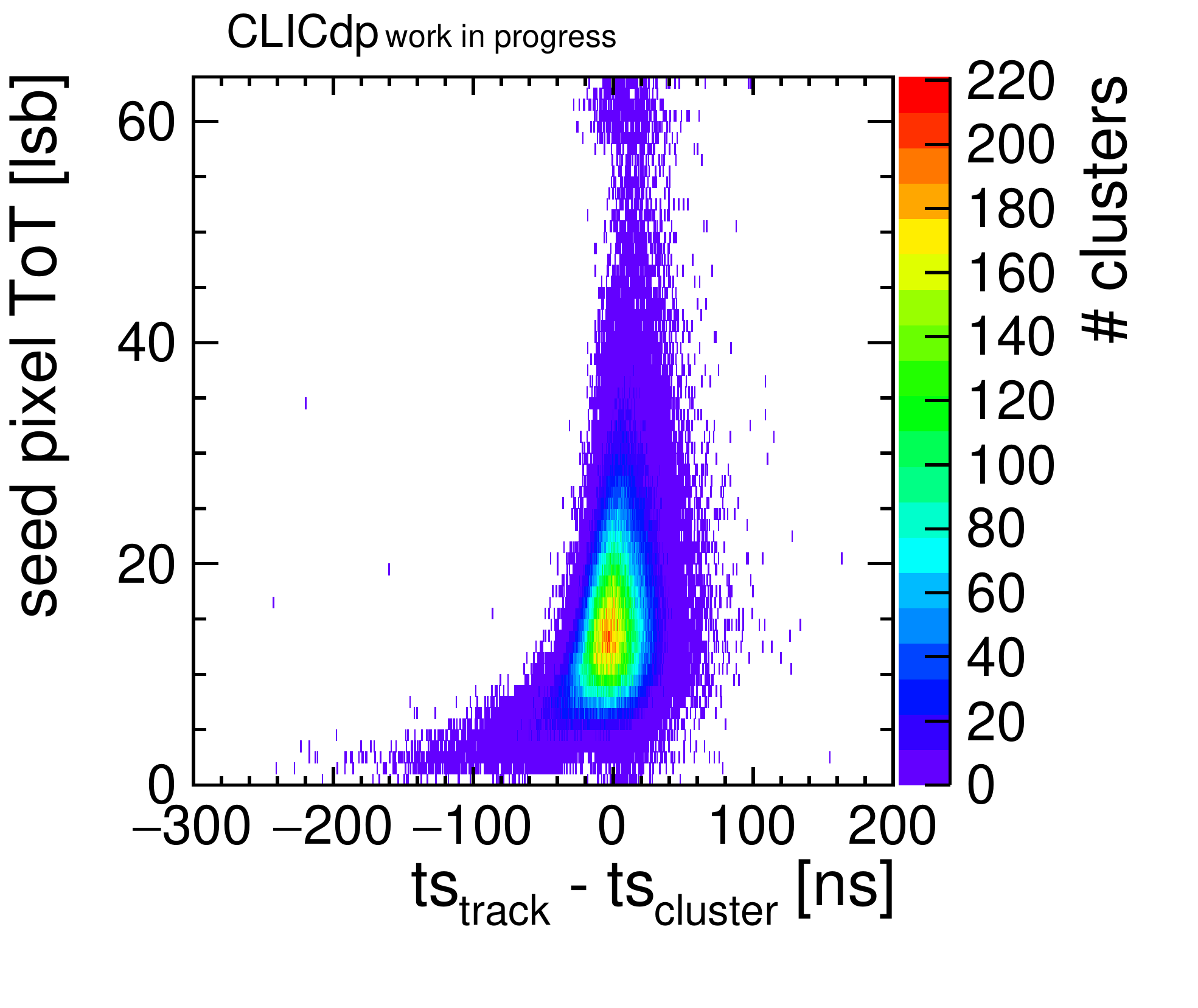}
\caption{\SI{80}{\ohm cm}, \SI{62}{\micro m}.}
\label{fig:apx_compare_timewalk_80}
\end{subfigure}%
\begin{subfigure}[b]{.33\linewidth}
\includegraphics[width=\textwidth]{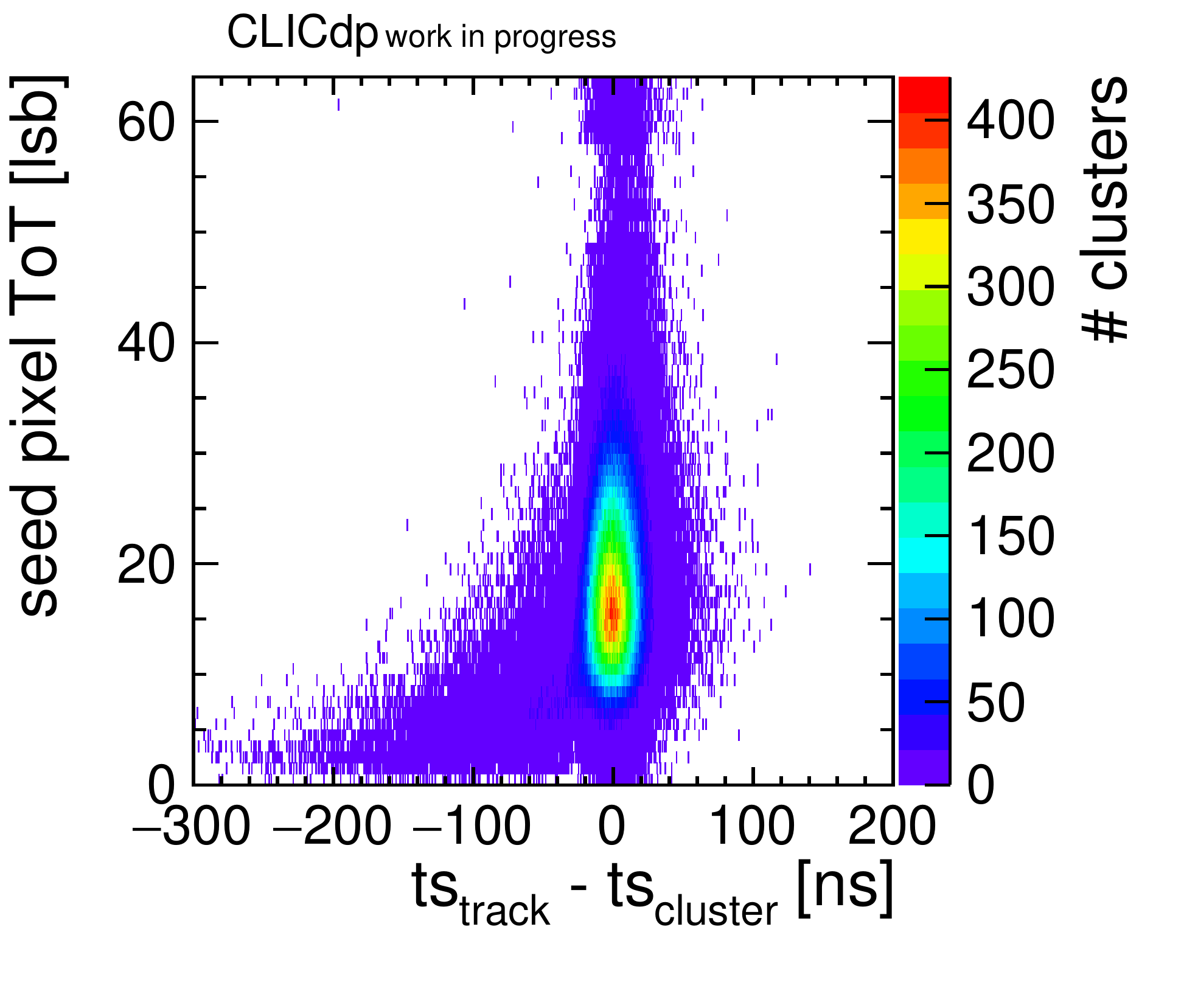}
\caption{\SI{200}{\ohm cm}, \SI{100}{\micro m}.}
\label{fig:apx_compare_timewalk_200}
\end{subfigure}
\caption{\label{fig:apx_compare_timewalk} Track-cluster time residual versus the seed pixel ToT of ATLASpix samples for different resistivities and thicknesses at a detection threshold of \SI{650}{e^-} and a bias voltage of \SI{-50}{V} before applying any correction.}
\end{figure}

Figure~\ref{fig:apx_compare_timewalk} shows 2D histograms of the seed pixel ToT of a cluster on the \textit{ATLASpix\_Simple} plotted against the time residual of the reference track and the associated cluster on the \textit{ATLASpix\_Simple}.
The reference track timestamp has been assigned by a Timepix3 plane~\cite{timepix3_paper} with a precision of \SI{1.56}{ns}~\cite{Pitters:2019yzg}.
The seed pixel of a cluster is the one with the earliest timestamp, which is used to define the timestamp of the whole cluster.
The ToT is a measure of the signal size, i.e.~a large ToT corresponds to a large signal size.
A clear dependence of the peak position of the time residual on the seed pixel ToT can be observed for the low substrate restivity of \SI{\sim 20}{\ohm cm} (see Figure~\ref{fig:apx_compare_timewalk_20}).
This effect, referred to as timewalk, is strongly reduced with increasing substrate resistivity (see Figures~\ref{fig:apx_compare_timewalk_80} and \ref{fig:apx_compare_timewalk_200}).

The dependence on the substrate resistivity can be explained as follows.
A higher substrate resistivity leads to a larger depleted volume for a given bias voltage.
As a consequence, more charge carriers can move freely and the induced signal is higher.
This means that the crossing of the detection threshold is steeper and less affected by jitter.
In addition, a larger electric field leads to a faster charge collection due to a shorter drift time of the charge carriers.
Hence, the timewalk effect is expected to be less pronounced.

\begin{figure}[b]
\centering
\includegraphics[width=0.45\textwidth]{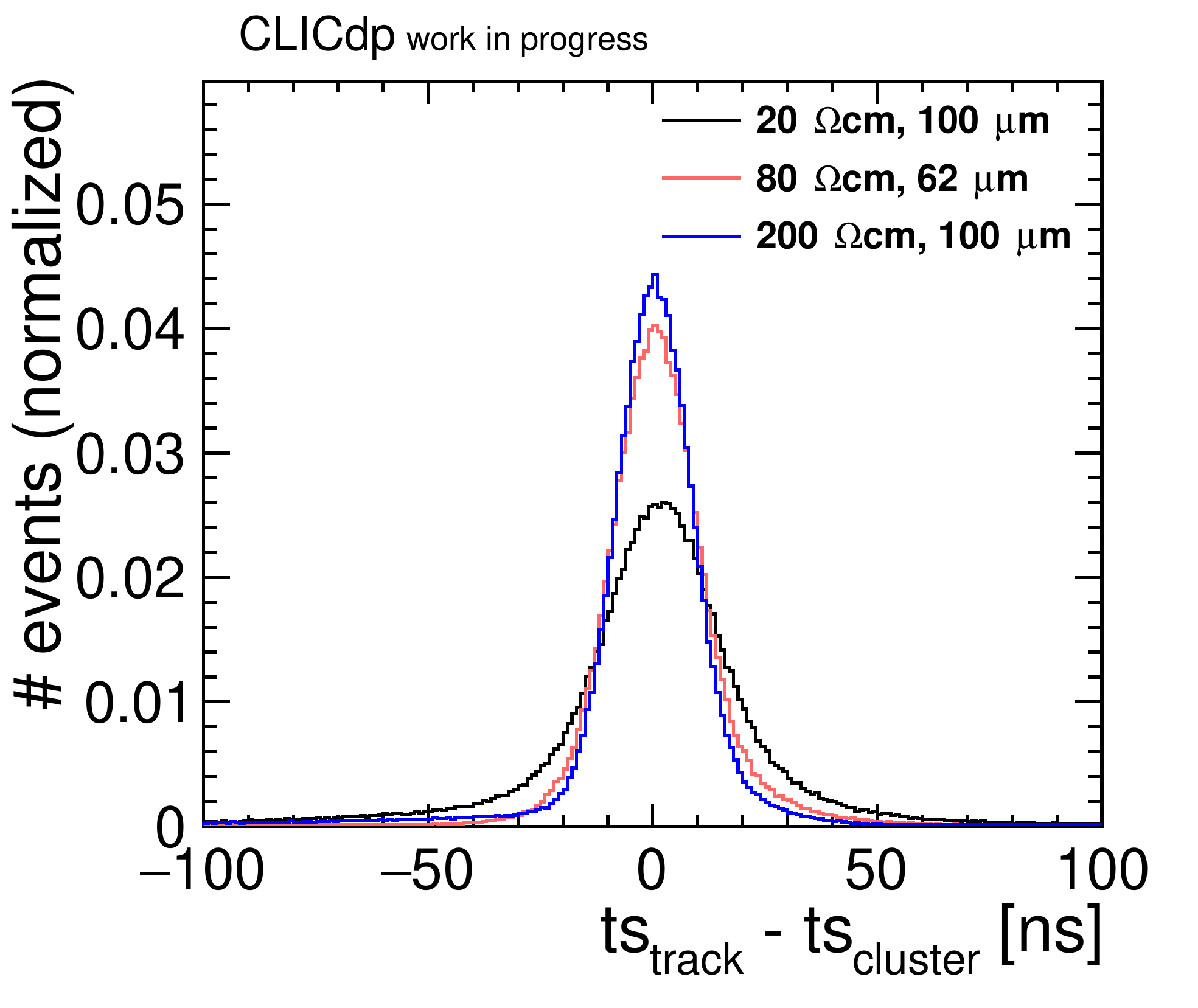}
\caption{\label{fig:apx_compare_time_residuals} Comparison of the track-cluster time residual for \textit{ATLASpix\_Simple} sensors with different substrate resistivities and thicknesses after row-dependent and timewalk correction.}
\end{figure}

A correction for a row-dependent signal delay as well as a timewalk correction of the cluster timestamp have been applied offline.
The resulting time residuals are shown in Figure~\ref{fig:apx_compare_time_residuals} for the three investigated substrate resistivities at equal operating conditions with a threshold of around \SI{650}{e^-} and a bias voltage of \SI{-50}{V}.
Gaussian fits have been performed to determine the time resolution as the standard deviation $\sigma$ of the fit function.
The results are summarized in Table~\ref{tab:apx_compare_time_resolutions}.

As in Figure~\ref{fig:apx_compare_timewalk}, a clear dependence on the substrate resistivity is observed and can be explained in the same way.
Hence, the timewalk effect is expected to be less pronounced and the timing resolution is better for a higher substrate resistivity.
The overall best result of $\sigma \sim 6.8$~\SI{}{ns} was obtained with a \SI{200}{\ohm cm} sample at a threshold of \SI{480}{e^-} and a bias voltage of \SI{-50}{V}.

It should be noted that all measured time residuals contain a contribution of about \SI{1.56}{ns} from the time resolution of the reference track timestamp.

\begin{table}[t]
\centering
\caption{Comparison of the time resolutions for \textit{ATLASpix\_Simple} samples with different substrate resistivities at a threshold of \SI{\sim 650}{e^-} and a bias voltage of \SI{-50}{V}. The quoted uncertainties correspond to the errors on the fit.}
\label{tab:apx_compare_time_resolutions}
\begin{tabular}{ccc}
\toprule
substrate resistivity [\SI{}{\ohm cm}] & thickness [\SI{}{\micro m}] & $\sigma$ [\SI{}{ns}] \\
\midrule
20 & 100 & $13.3\pm0.04$ \\
80 & 62 & $9.4\pm0.03$ \\
200 & 100 & $8.3\pm0.02$ \\
\bottomrule
\end{tabular}
\end{table}

\subsection{The CLICTD HR-CMOS Prototype}
\label{subsec:clictd}
\begin{figure}[b]
\centering
\includegraphics[width=0.35\textwidth]{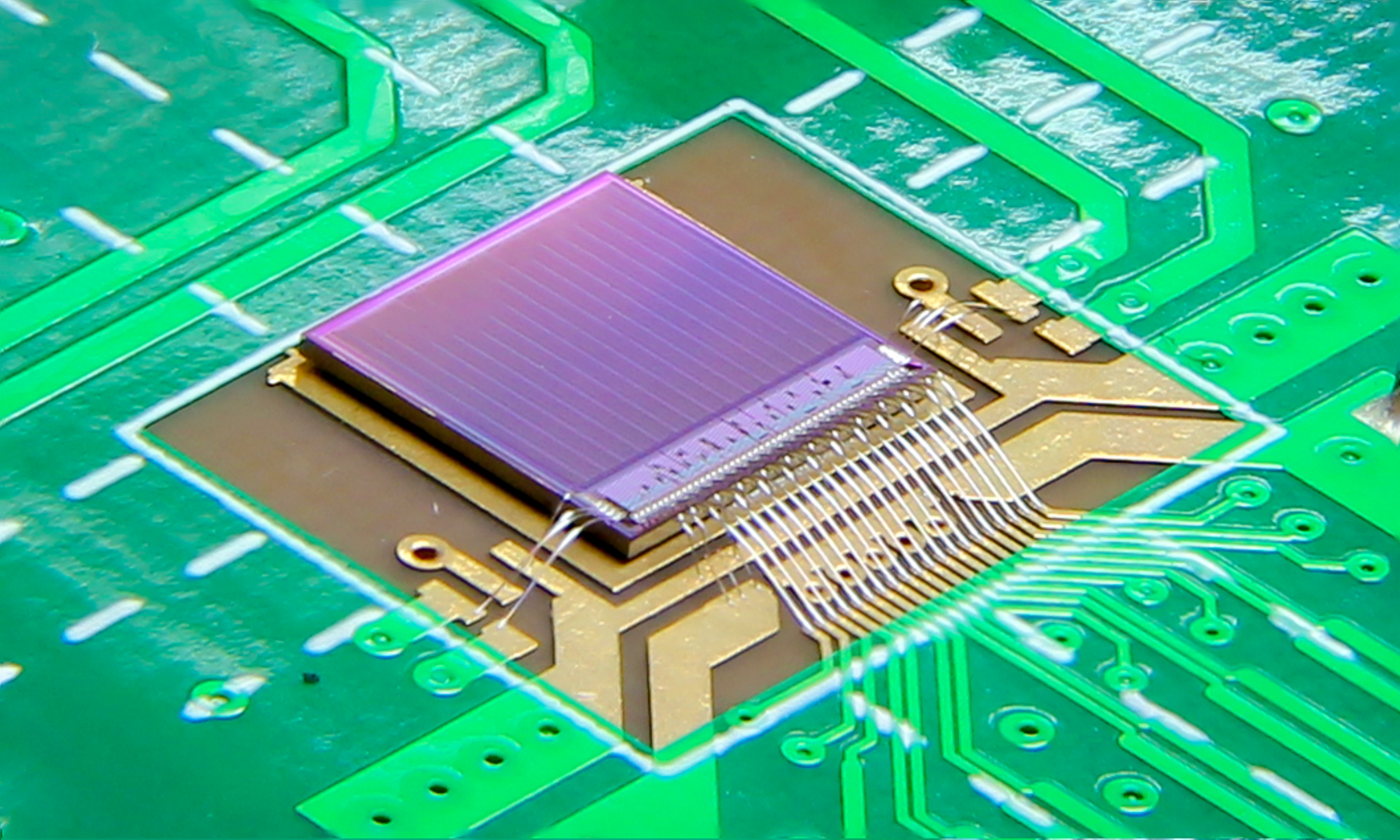}
\caption{Photograph of a \textit{CLICTD} sensor glued and wire-bonded to a printed circuit board.}
\label{fig:clictd_picture}
\end{figure}

The \textit{CLICTD} sensor has been designed to meet the requirements of the \textit{CLIC} tracking detector.
It is fabricated in a modified CMOS imaging process, which has been optimised with the help of 3D TCAD studies~\cite{Munker_2019}.
The chip features a matrix of $128\times128$ pixels with a pitch of $37.5\times30$~\SI{}{\micro m\squared}.
A photograph is shown in Figure~\ref{fig:clictd_picture}.
In an innovative readout segmentation scheme, $8\times1$ pixels each are combined into a single readout channel with a combined 8-bit ToA measurement with \SI{10}{ns} binning and a 5-bit ToT measurement.
This scheme allows to save space for digital circuitry while maintaining the small-collection electrode design at a low pixel pitch~\cite{Kremastiotis:2019pnb}.

The sensor has been characterised both in laboratory and test-beam measurement campaigns.

\subsubsection{Hit Detection Efficiency}

\begin{figure}[t]
\centering
\begin{subfigure}[b]{.45\linewidth}
\centering
\includegraphics[width=\textwidth]{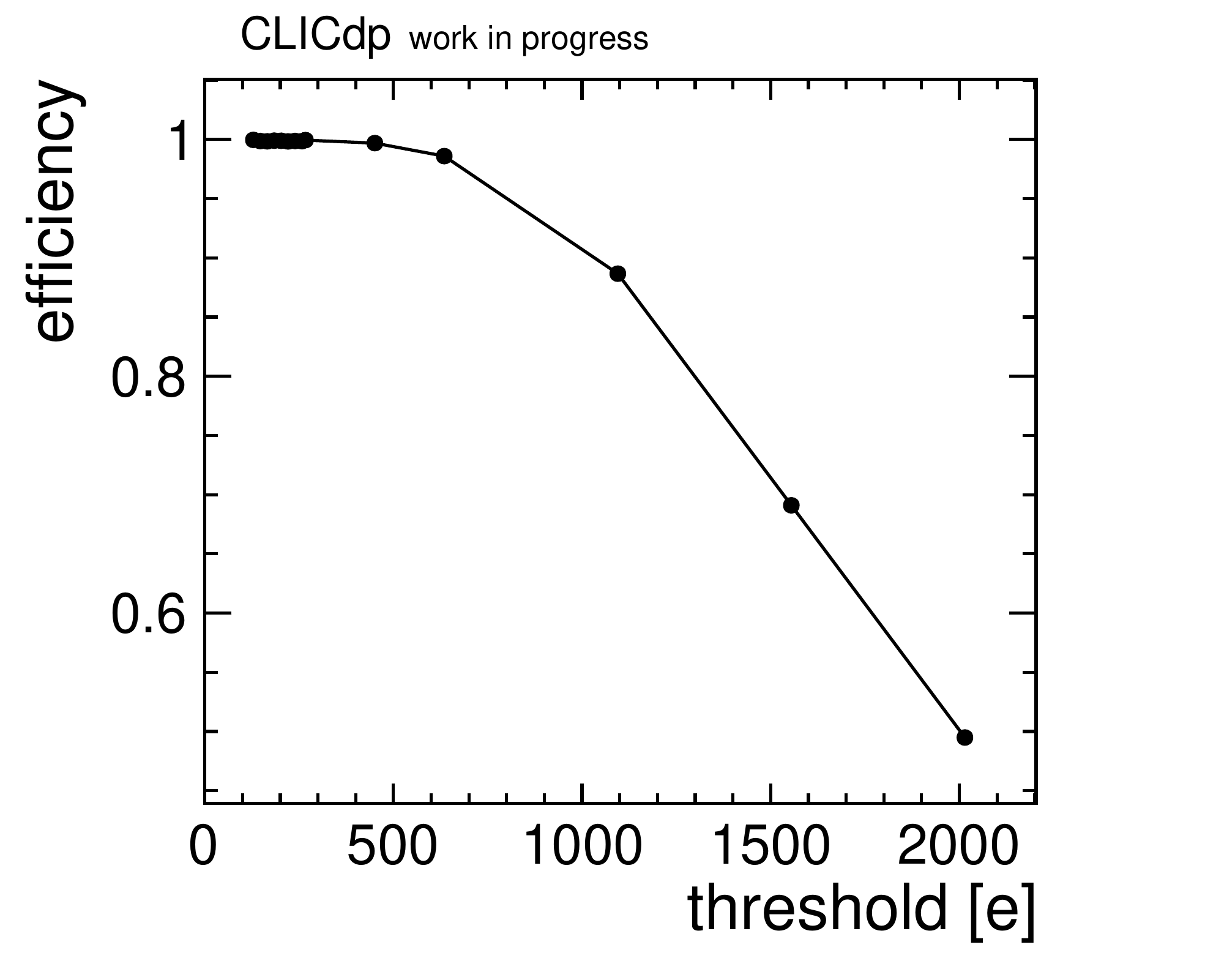}
\caption{Hit detection efficiency for a threshold range from 150 to \SI{2500}{e^-}.}
\label{fig:clictd_efficiency_thres}
\end{subfigure}%
\hfill
\begin{subfigure}[b]{.45\linewidth}
\centering
\includegraphics[width=\textwidth]{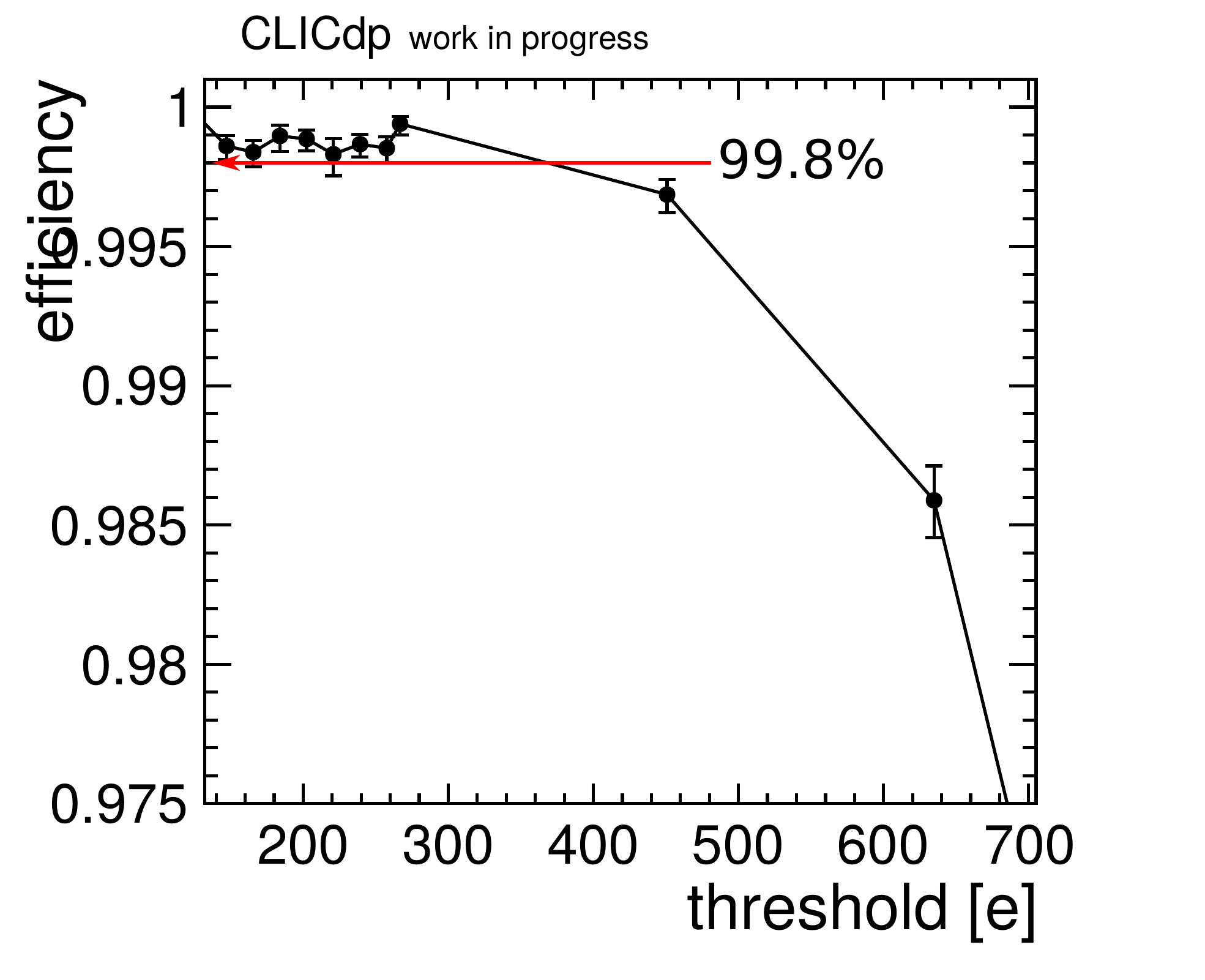}
\caption{Hit detection efficiency for a threshold range from 150 to \SI{700}{e^-}.}
\label{fig:clictd_efficiency_thres_zoom}
\end{subfigure}
\caption{\label{fig:clictd_efficiency} Hit detection efficiency measurements for a \textit{CLICTD} sample.}
\end{figure}

\begin{figure}[b]
\centering
\includegraphics[width=0.45\textwidth]{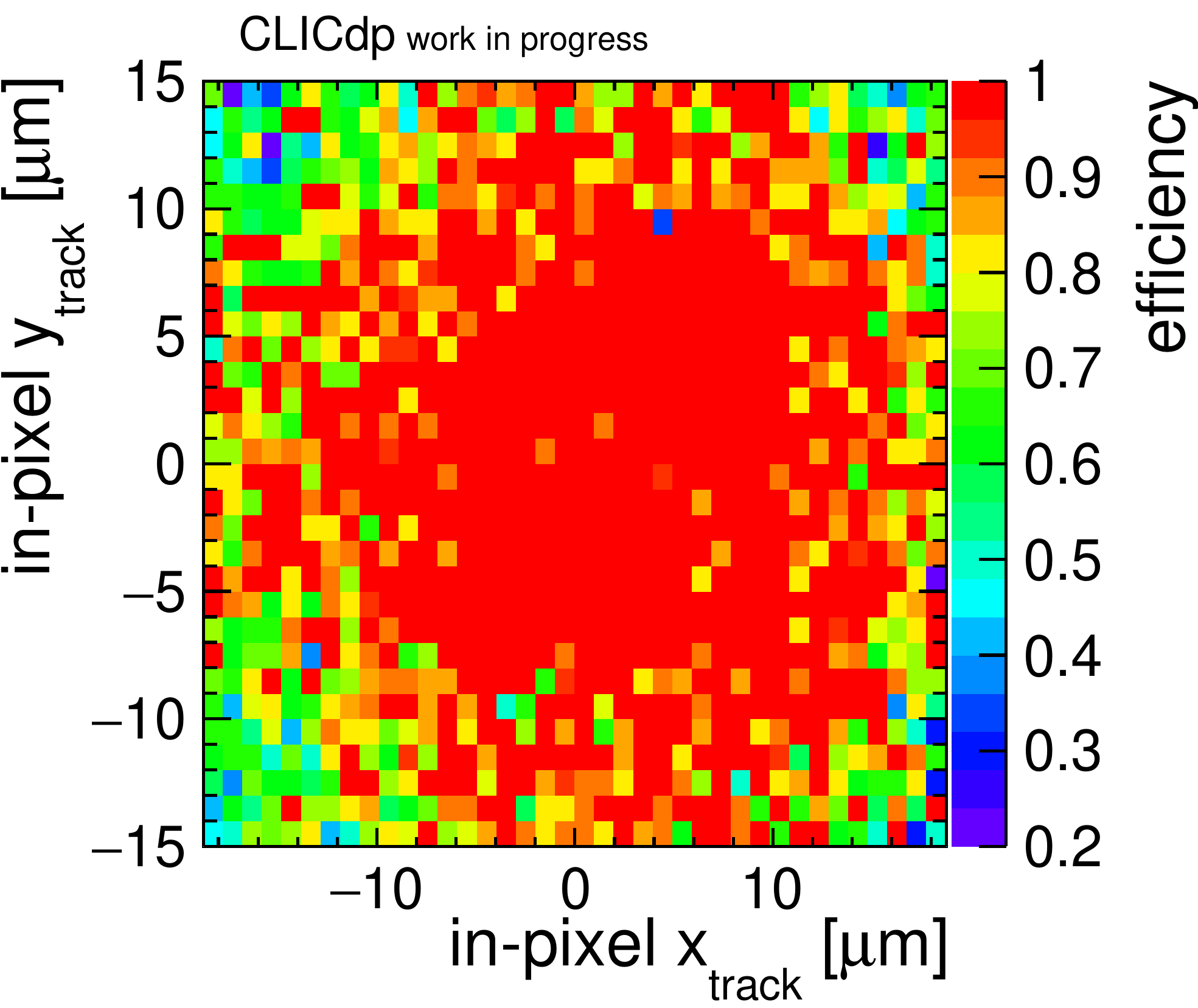}
\caption{In-pixel hit detection efficiency of a \textit{CLICTD} sample for a threshold of \SI{1500}{e^-}.}
\label{fig:clictd_efficiency_inpixel}
\end{figure}

Measurements at the DESY-II test-beam facility~\cite{desy2_testbeam_facility} show that the \textit{CLICTD} sensor can be operated fully efficiently, i.e.~with a hit detection efficiency above \SI{99.8}{\percent} for threshold values below \SI{400}{e^-}, as can be seen in Figures~\ref{fig:clictd_efficiency_thres} and \ref{fig:clictd_efficiency_thres_zoom}.
It has been measured down to a threshold of \SI{150}{e^-}, where the sensor can be operated without noise hits.
At thresholds above \SI{400}{e^-}, the hit detection efficiency drops because the induced signal of the collected charge is less likely to cross the detection threshold.

As the in-pixel distribution of the hit detection efficiency in Figure~\ref{fig:clictd_efficiency_inpixel} reveals, the efficiency remains high in the centre of a pixel and drops towards edges and especially the corners.
This is explained by an increased amount of charge sharing into the neighbouring cells for particles incident close to the edges and corners of the sensor.
As a consequence, the collected charge per pixel is smaller and the induced signal is less likely to cross the detection threshold.

\subsubsection{Timing Performance}
The time resolution of the \textit{CLICTD} sensor can be deduced from the width of the time residual between the track timestamp and the \textit{CLICTD} cluster timestamp, which is defined as the earliest pixel timestamp within the cluster.

\begin{figure}[t]
\centering
\begin{subfigure}[b]{.45\linewidth}
\centering
\includegraphics[width=\textwidth]{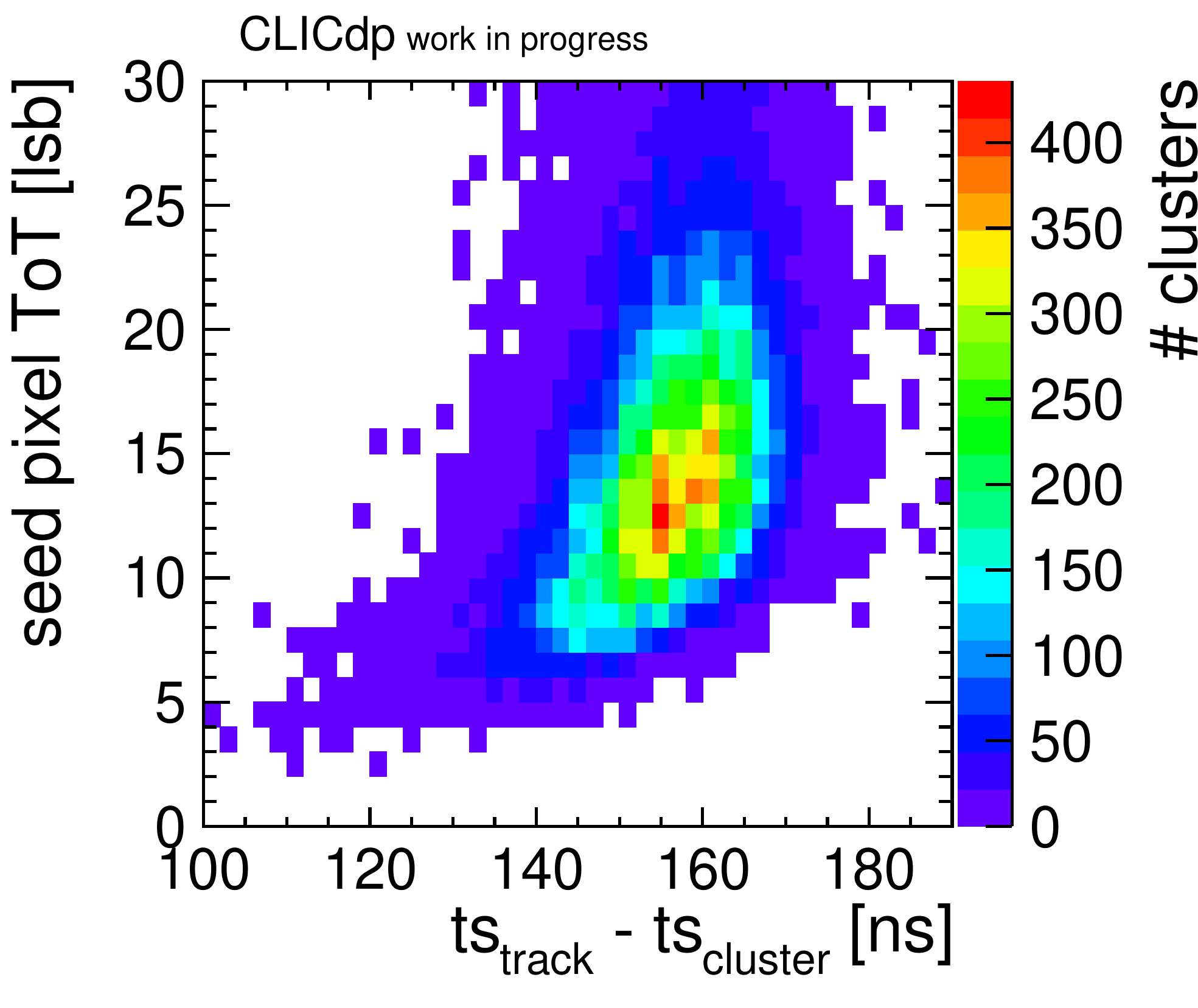}
\caption{Track-cluster time residual vs.~seed pixel ToT of the \textit{CLICTD}.}
\label{fig:clictd_timewalk}
\end{subfigure}%
\hfill
\begin{subfigure}[b]{.45\linewidth}
\centering
\includegraphics[width=\textwidth]{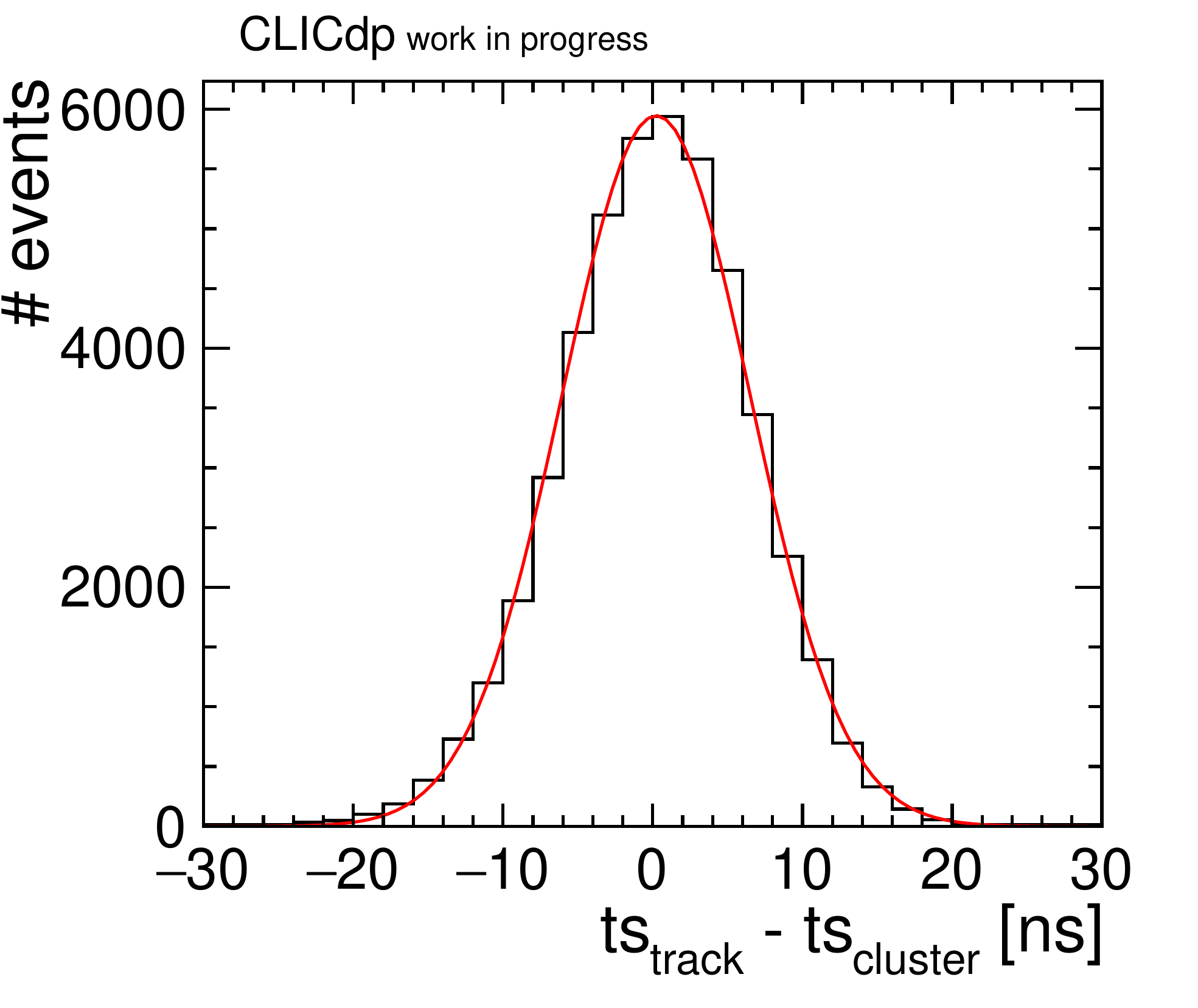}
\caption{Track-cluster time residual after timewalk correction.}
\label{fig:clictd_timeresidual}
\end{subfigure}
\caption{\label{fig:clictd_timing} Timing performance measurements for the \textit{CLICTD}.}
\end{figure}

Figure~\ref{fig:clictd_timewalk}, which shows the track-cluster time residual plotted against the seed pixel ToT of the cluster, reveals a clear timewalk effect.
A timewalk correction can be applied offline to improve the time resolution.
The result is shown in Figure~\ref{fig:clictd_timeresidual}.
A Gaussian fit to the time residuals results in a time resolution of $\sigma \sim 6.3$~\SI{}{ns}.

\section{Summary and Outlook}
\label{sec:summary_and_outlook}
The experimental conditions at the proposed linear $e^+e^-$ collider \textit{CLIC} pose stringent requirements on the detector performance.
To meet these requirements, an all-silicon vertex and tracking detector system is foreseen, for which a diverse R\&D programme is being pursued in order to identify suitable technologies.
As part of these efforts, hardware and software tools for simulation, data acquisition, as well as test-beam analysis have been developed within the CLICdp collaboration.

For the ultra-low mass vertex detector, different small pitch hybrid technologies with innovative interconnection techniques are under investigation.
For the large-scale tracking detector, the focus of the R\&D lies on depleted monolithic CMOS technologies.
Various sensor prototypes, such as the \textit{ATLASpix\_Simple} and the \textit{CLICTD}, are being characterized in laboratory and test-beam measurement campaigns.

Samples of the \textit{ATLASpix\_Simple} HV-MAPS test chip with different substrate resistivies have been compared in view of the timing performance as well as the hit detection efficiency.
The \textit{ATLASpix\_Simple} reaches hit detection efficiencies above \SI{99.8}{\percent} and a timing resolution of \SI{\sim 6.8}{ns}.
A high substrate resistivity is essential both for an excellent timing behaviour as well as a larger operating window maintaining a high hit detection efficiency.

The \textit{CLICTD} HR-CMOS sensor prototype has been designed to explore a novel readout segmentation scheme.
A hit detection efficiency above \SI{99.8}{\percent} and a time resolution as good as \SI{\sim 6.3}{ns} have been measured.

In summary, the presented results show that HV-MAPS and HR-CMOS are promising technologies and have a large potential to be used in the \textit{CLIC} tracking detector.
Most of the detector requirements are fulfilled by the test sensors and further prototypes with newer process technologies and improved designs are planned to be fabricated and tested in the future.

\acknowledgments

This work has been sponsored by the Wolfgang Gentner Programme of the German Federal Ministry of Education and Research (grant no. 05E15CHA).

The measurements leading to these results have been performed at the Test Beam Facility at DESY Hamburg (Germany), a member of the Helmholtz Association (HGF).

We would like to gratefully acknowledge CERN and their accelerator staff for the reliable test-beam operation.
This project has received funding from the European Union's Horizon 2020 research and innovation programme under grant agreement No 654168.
It benefited from services provided by the ILC Virtual Organisation, supported by the national resource providers of the EGI Federation and was done using resources provided by the Open Science Grid, which is supported
by the National Science Foundation and the U.S. Department of Energy's Office of Science.

\bibliography{references}



\end{document}